\documentclass [prl,twocolumn,longbibliography]{revtex4-2}

\usepackage{graphicx}
\usepackage{subfigure}
\usepackage{amsmath}
\usepackage{amsthm}
\usepackage{amssymb}
\usepackage{hyperref}
\usepackage{xcolor}
\usepackage{textcomp}
\usepackage[normalem]{ulem}

\hypersetup{colorlinks=true, linkcolor=black, citecolor=black, urlcolor=blue}

\newcommand{\ket}[1]{\left| #1 \right>} % for Dirac bras
\newcommand{\braket}[2]{\left< #1 \vphantom{#2} \right| \left. #2 \vphantom{#1} \right>} % for Dirac brackets

\makeatletter
\newcommand{\widefigcaption}[1]{%
  \refstepcounter{figure}%
  \@makecaption{\fnum@figure}{#1}\par
}
\makeatother

\begin{document}

%=========================================================================
\title{Driven-Dissipative Landau Polaritons: Two Highly Nonlinearly-Coupled Quantum Harmonic Oscillators}
 
\author{Farokh Mivehvar}
\email[Corresponding author: ]{farokh.mivehvar@uibk.ac.at}
\affiliation{Institut f\"ur Theoretische Physik, Universit{\"a}t Innsbruck, A-6020~Innsbruck, Austria}
\begin{abstract}
Landau levels (LLs) are the massively-degenerate discrete energy spectrum of a charged particle in a transverse magnetic field and lie at the heart of many intriguing phenomena such as the integer and fractional quantum Hall effects as well as quantized vortices. In this Letter, we consider coupling of LLs of a transversely driven, single charge-neutral particle in a synthetic gauge potential to a quantized field of an optical cavity---a setting reminiscent of superradiant self-ordering setups in quantum gases. We uncover that this complex system can be surprisingly described in terms of two highly nonlinearly-coupled quantum harmonic oscillators, thus enabling a full quantum mechanical treatment. Light-matter coupling mixes the LLs and the superradiant photonic mode, leading to the formation of hybrid states referred to as ``Landau polaritons''. They inherit partially the degeneracy of the LLs and possess intriguing features such as non-zero light-matter entanglement and quadrature squeezing. Depending on the system parameters and the choice of initial state, the system exhibits diverse nonequilibrium quantum dynamics and multiple steady states, with distinct physical properties. This work lays the foundation for further investigating the novel, driven-dissipative Landau-polariton physics in quantum-gas--cavity-QED settings. 
\end{abstract}

\maketitle

%=========================================================================
\emph{Introduction.}---In quantum mechanics, the energy spectrum of two-dimensional electrons pierced by a uniform magnetic field are quantized into discrete, equally spaced levels, known as Landau levels (LLs). LLs are massively degenerate and possess very distinct electronic and optical properties, giving rise to intriguing phenomena such as quantized conductance and robust edge transport in quantum Hall materials~\cite{PrangeGirvin1990}. 

The interaction between quantum Hall materials and quantum fluctuations of cavity \emph{vacuum} electromagnetic fields has recently gained significant attention~\cite{Scalari2012, Zhang2016Collective, Schlawin2022}. In particular, it has been demonstrated vacuum fluctuations weaken the topological protection of integer quantum Hall states, due to the \emph{virtual}-photon--mediated long-range electron hopping and the emergence of a soft polariton mode~\cite{Keller2020Landau, Rokaj2023Weakened}, leading to the development of a finite resistivity~\cite{Appugliese2022}. In the case of fractional quantum Hall states interacting with vacuum fluctuations~\cite{Enkner2025}, it has been predicted that beyond the dipole approximation Kohn’s theorem breaks down, leading to the hybridization of matter and light excitations and the formation of plasmon polaritons~\cite{Winter2025Fractional} and graviton polaritons~\cite{Bacciconi2025Theory}.   

\begin{figure}[t!]
\centering
\includegraphics [width=0.49\textwidth]{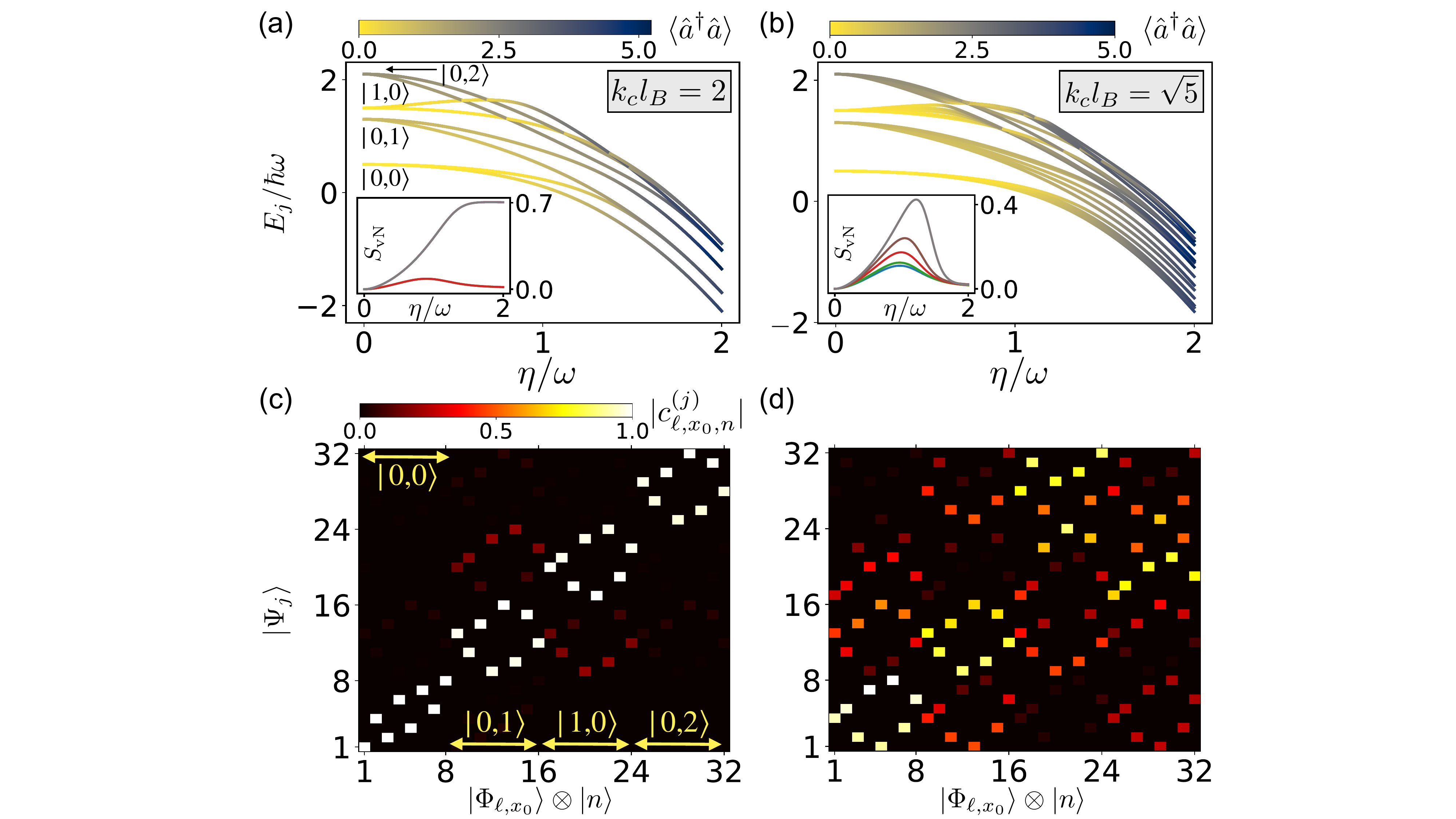}
\caption{The lowest four energy bands $E_j$ as a function of the light-matter coupling $\eta$ for two different values of $k_cl_B=2$ (a) and $\sqrt{5}$ (b). In the zero-coupling limit in panel (a), 
the uncoupled LLs and the photon Fock states $\ket{\Phi_{\ell,x_0}}\otimes\ket{n}$ are indicated by a short form as $\ket{\ell,n}$ with the $x_0$ degeneracy implied. The color map of the energy bands displays the average photon $\langle \hat{a}^\dag\hat{a}\rangle$ content of the corresponding states. The insets show the von Neumann entropy $S_{\rm vN}$ of the reduced system as a function of $\eta$ for the lowest band (composed of eight sub-bands). Absolute value of the overlap $|c_{\ell,x_0,n}^{(j)}|$ of the Landau polaritons $\ket{\Psi_j}$ with the uncoupled states for $k_cl_B = \sqrt{5}$ and two different coupling strengths $\eta/\omega=0.1$ (c) and $1$ (d). The order of the uncoupled states is indicated in panel (c). For all the plots, $\Delta_c=-0.8\omega$ and $x_0=\pi l_B\{-1,-3/4,-1/2,\cdots,3/4\}$.} 
\label{fig:energy_spectra}
\end{figure}

\begin{figure}[t!]
\centering
\includegraphics [width=0.49\textwidth]{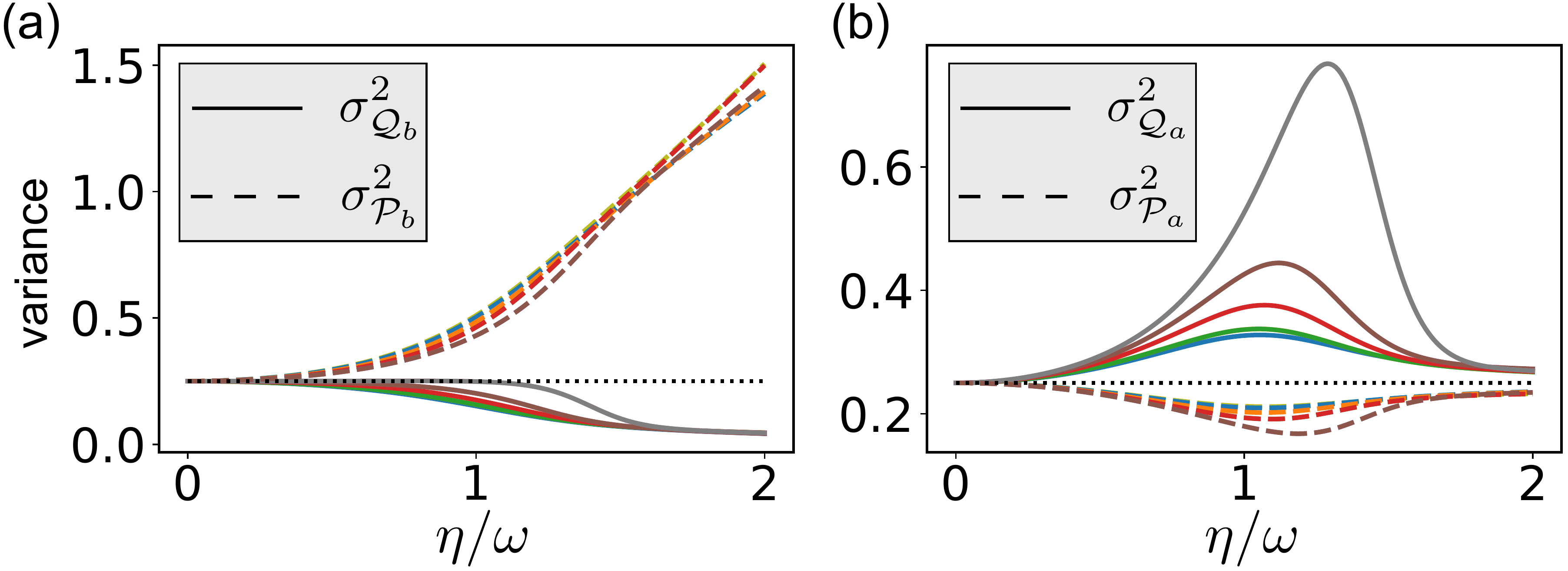}
\caption{Variance $\{\sigma^2_{{\mathcal Q}_{O}}, \sigma^2_{{\mathcal P}_{O}} \}$ of the atomic (a) and photonic~(b) quadrature operators in the lowest band (consisting of eight sub-bands) as a function of the atom-photon coupling strength $\eta$ for $k_cl_B=\sqrt{5}$. While for the Landau oscillator the position quadrature is squeezed, $\sigma^2_{{\mathcal Q}_{b}}<1/4$ (i.e., below the dotted line), for the photonic oscillator the momentum quadrature is squeezed, $\sigma^2_{{\mathcal P}_{a}}<1/4$. 
}
\label{fig:squeezing}
\end{figure}

\begin{figure*}[t!]
\centering
\includegraphics [width=0.97\textwidth]{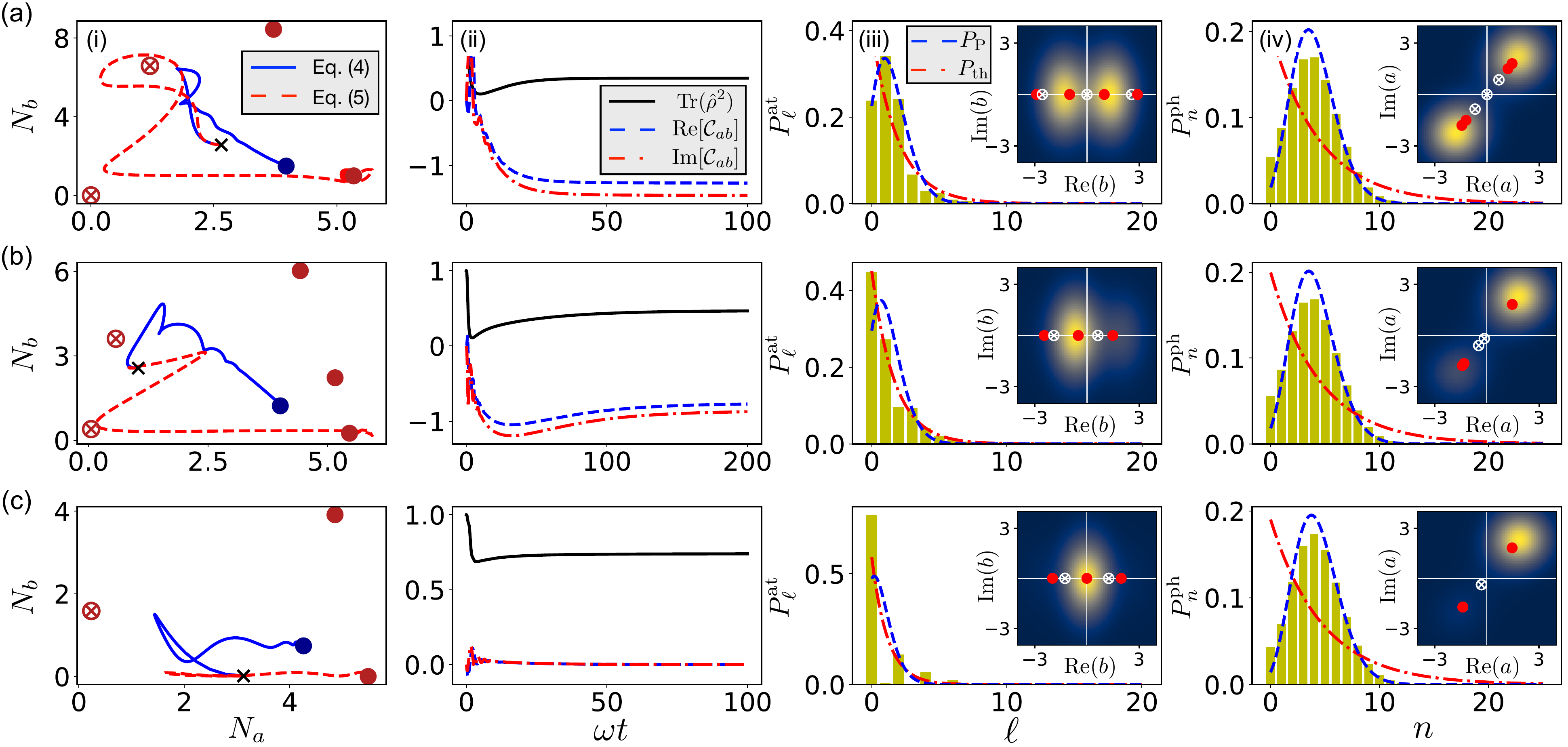}
\caption{The nonequilibrium dynamics of the system for $\eta=3\omega$, $\kappa=\omega$, $k_cl_B=1$, and three different values of $x_0/l_B=-\pi/2$~(a), $-3\pi/4$ (b), and $-\pi$ (c). The column (i) shows the quantum (solid blue) and semi-classical (dashed red) dynamics of the system in the phase space of the average photon vs.\ average LL-occupation number. The filled (crossed) dark-red circles represent the stable (unstable) semi-classical fixed points, while the filled dark-blue circles are the quantum steady states. The black crosses are the chosen initial product coherent states: $\ket{1.6}\otimes\ket{1.6-0.15i}$ (a), $\ket{1.6+0.05i}\otimes\ket{1+0.2i}$ (b), and $\ket{0.1+0.05i}\otimes\ket{1.75-0.25i}$ (c). The column (ii) shows the lowest order correlation $\mathcal{C}_{ab}$ between the two oscillators as well as the statistical mixing $\text{Tr}(\hat{\rho}^2)$. The steady-state atomic and photonic distributions are depicted in the columns (iii) and (iv), respectively, with the Poisson $P_{\rm P}$ and thermal $P_{\rm th}$ distributions represented as references. The corresponding steady-state atomic and photonic $Q$ functions are shown in the insets, with the filled red (crossed white) circles representing the semi-classical stable (unstable) steady states. 
} 
\label{fig:dynamics}
\end{figure*}

In this Letter, we go beyond the effects of quantum fluctuations of cavity vacuum fields in the LLs~\cite{Nataf2019Rashba}, by considering a genuine \emph{driven-dissipative} scenario which allows \emph{position-dependent} hybridization of the LLs~\cite{Fletcher2021} and \emph{real} cavity photons. Our proposed setup consists of a transversely-driven single atom subject to a synthetic magnetic field inside an off-resonant, single-mode optical cavity---a setting reminiscent of superradiant self-ordering setups in quantum-gas--cavity-QED systems~\cite{Mivehvar2021}. We uncover that this complex system is equivalent to two highly nonlinearly-coupled harmonic oscillators, leading to a significant reduction of the Hilbert-space dimension required for the full quantum-mechanical description of relevant low-energy physics. Due to the absence of no-go conditions here~\cite{Vukics2012Adequacy}, the light-matter coupling results in superradiant-type photon scattering into the cavity, in sharp contrast to solid-state cavity quantum-Hall materials~\cite{Scalari2012, Zhang2016Collective, Schlawin2022, Keller2020Landau, Rokaj2023Weakened, Appugliese2022, Enkner2025, Winter2025Fractional, Bacciconi2025Theory}. As a consequence, the LLs are strongly mixed with these real cavity photons as shown in Fig.~\ref{fig:energy_spectra}, leading to the formation of hybrid quasiparticles referred to as ``\emph{Landau polaritons}''. These states inherit part of the LLs' degeneracy and exhibit remarkable properties such as non-zero light-matter entanglement and quadrature squeezing; see Fig.~\ref{fig:squeezing}. Depending on the system parameters and the choice of initial state, the quantum dynamics display a rich variety of nonequilibrium behaviors as depicted in Fig.~\ref{fig:dynamics} and can support multiple steady states with distinct physical characteristics. The many-body topological aspects and implications of this system can be of great importance to broad physics communities and will be investigated in future works.

%=========================================================================
\emph{System.}---Consider a single atom confined in a box trap within the $x$-$y$ plane and subject to an external uniform synthetic magnetic field $\boldsymbol{\mathcal{B}}=\mathcal{B}\mathbf{e}_z=\boldsymbol{\nabla}\times\boldsymbol{\mathcal A}(\hat{\mathbf r})$ along the $z$ axis. The box potential is located inside an optical cavity aligned along the $x$ axis, as illustrated in Fig.~\ref{EM-fig:setup} in End Matter (EM). The atom couples to a single mode of the cavity with the coupling strength $\mathcal{G}(\hat{x})=\mathcal{G}_0\cos(k_c \hat{x})$, where $k_c=2\pi/\lambda_c=\omega_c/c$ is the wavenumber of the cavity mode. Additionally, the atom is driven along the $z$ axis by a transverse pump laser with frequency $\omega_p$ and amplitude $\Omega_0$. Both the cavity and pump fields are far detuned from all atomic electronic transitions, but closely detuned with respect to each other. This ensures that the atom remains in its electronic ground state, while low-lying external states can be excited due to two-photon Raman processes.

In this dispersive limit and in the rotating frame of the pump laser, the system is described by the Hamiltonian~\cite{Masalaeva2025Rotational},
\begin{align} \label{eq:2D_L_H}
\hat{H}&=\frac{1}{2M}\left[\hat{\mathbf{p}} - \boldsymbol{\mathcal{A}}(\hat{\mathbf r})\right]^2
-\hbar\Delta_c\hat{a}^\dag\hat{a}
+\hbar\eta(\hat{a}^\dag+\hat{a})\cos(k_c \hat{x}),
\end{align}
where $M$ is the atomic mass, $\{\hat{a},\hat{a}^\dag\}$ the cavity photon operators, and $\Delta_c\equiv\omega_p-\omega_c$  the pump-cavity detuning. The synthetic, time-independent vector potential $\boldsymbol{\mathcal{A}}$ is minimally coupled to the momentum operator $\hat{\mathbf{p}}$. The atom-photon coupling due to the two-photon Raman processes is encoded in $\eta=\mathcal{G}_0\Omega_0/\Delta_a$ (with $\Delta_a$ being the detuning of the pump laser from the closest atomic excited state). Note that the synthetic charge is set to unity for simplicity and the optomechanical term, which is insignificant in many current experiments~\cite{Mivehvar2021}, has been omitted in Eq.~\eqref{eq:2D_L_H}.

Since physical observables such as the energy spectrum must be independent from the choice of a specific gauge for $\boldsymbol{\mathcal A}(\hat{\mathbf r})$~\cite{Swenson1989The}, without loss of generality I choose the Landau gauge along the $y$ direction, $\boldsymbol{\mathcal A}(\hat{\mathbf r})=\mathcal{A}_y(\hat{\mathbf r})\mathbf{e}_y = \mathcal{B} \hat{x}\mathbf{e}_y$. This renders the system translationally invariant along the $y$ direction and the Hamiltonian~\eqref{eq:2D_L_H} of the system can, therefore, be simplified and recast in one spatial dimension as,
\begin{align} \label{eq:1D_L_H}
\hat{H}&=\frac{\hat{p}_x^2}{2M}+\frac{1}{2}M\omega^2(\hat{x}-x_0)^2
-\hbar\Delta_c\hat{a}^\dag\hat{a}\nonumber
\\
&+\hbar\eta(\hat{a}^\dag+\hat{a})\cos(k_c \hat{x}),
\end{align}
where $\omega \equiv \mathcal{B}/M$ is the cyclotron frequency and $x_0 \equiv \hbar k_y/\mathcal{B}$, with $\hbar k_y$ being the eigenvalues of the operator~$\hat{p}_y$.

The free atomic part of the Hamiltonian~\eqref{eq:1D_L_H}, $\hat{H}_a\equiv \hat{p}_x^2/{2M}+M\omega^2(\hat{x}-x_0)^2/2$, is just a shifted harmonic oscillator with well-known harmonic oscillator wavefunctions $\braket{x}{\phi_{\ell,x_0}}=\phi_{\ell}(x-x_0)$ and energies $\epsilon_{\ell,x_0}=\hbar\omega(\ell+1/2)$~\cite{LandauLifshitz1977}. Note that the LLs $\braket{x,y}{\Phi_{\ell,x_0}}=\Phi_{\ell,x_0}(x,y)=e^{i\mathcal{B}x_0y/\hbar}\phi_{\ell}(x-x_0)$ for a given $\ell$ are highly degenerate as the energy spectrum $\epsilon_{\ell,x_0}$ does not depend on $x_0$.  

%=========================================================================
\emph{Coupled oscillator model.}---By noting that the free cavity Hamiltonian $\hat{H}_c\equiv-\hbar\Delta_c\hat{a}^\dag\hat{a}$ is also obviously a quantum harmonic oscillator, Hamiltonian~\eqref{eq:1D_L_H} can be interpreted as two coupled harmonic oscillators. This can be made explicit by introducing the shifted matter ladder operator $\hat{b}\equiv[(\hat{x}-x_0)/l_B+il_B\hat{p}_x/\hbar]/\sqrt{2}$ and its Hermitian conjugate $\hat{b}^\dag$, where $\hat{b}$ and $\hat{b}^\dag$ follow the bosonic commutation relation $[\hat{b},\hat{b}^\dag]=1$ and $l_B\equiv\sqrt{\hbar/M\omega}=\sqrt{\hbar/\mathcal{B}}$ is the magnetic length. The Hamiltonian~\eqref{eq:1D_L_H} then takes the following revealing two-mode form,
\begin{align} \label{eq:coupled_QHO}
\hat{H}&=-\hbar\Delta_c\hat{a}^\dag\hat{a}
+\hbar\omega\left(\hat{b}^\dag\hat{b}+\frac{1}{2}\right)\nonumber\\
&+\hbar\eta(\hat{a}^\dag+\hat{a})
\cos\left[k_cl_B \left( \frac{\hat{b}^\dag+\hat{b}}{\sqrt{2}} + \frac{x_0}{l_B} \right) \right].
\end{align}
This coupled quantum harmonic oscillator model~\eqref{eq:coupled_QHO} is unique in numerous aspects and is the central finding of this work. 

Let us look closely at the Hamiltonian~\eqref{eq:coupled_QHO}. First and foremost, I note that expanding the cosine using its power series, $\cos(\theta)=\sum_{m=0}^\infty (-1)^m \theta^{2m}/(2m)!$, yields infinite-order couplings between the matter $\hat{b}$ and field $\hat{a}$ quantum oscillators. The strength of these couplings is determined not only by the atom-photon coupling $\eta$, but also by the ratio $k_cl_B=2\pi l_B/\lambda_c$ between the magnetic length $l_B$ and the cavity wavelength $\lambda_c$. Only for very small ratios $k_cl_B\ll1$, one can truncate the series expansion of the cosine and obtain an explicit finite-order nonlinear model, in a close analogy to the Lamb--Dicke regime in ion physics~\cite{Wineland1998}. Note also that the guiding center $x_0$ now appears explicitly in the coupling term, signaling that the degeneracy of the LLs should be lifted, at least partially, by the atom-photon coupling. Furthermore, when $k_cx_0=(2j+1)\pi/2$ with $j\in \mathbb{Z}$, the Hamiltonian~\eqref{eq:coupled_QHO} becomes invariant under the parity transformation $\hat{a}\to-\hat{a}$ and $\hat{b}\to-\hat{b}$, acquiring therefore a $\mathbb{Z}_2$ symmetry. This has important consequences in the quantum dynamics of the system as we will see later.

%=========================================================================
\emph{Landau polaritons.}---The Hamiltonian~\eqref{eq:coupled_QHO} has five independent free parameters $\{\omega, \Delta_c, \eta, k_c, x_0 \}$ (recall that $\Delta_c$ is the pump-cavity detuning and can be tuned independently of $k_c$; for simplicity, we choose $M=1$ and therefore $\omega=\mathcal{B}$). Diagonalizing the Hamiltonian as a function of the atom-photon coupling strength~$\eta$ yields the energy spectrum $E_j(\eta)$ and the corresponding eigenstates $\ket{\Psi_j}$ of the system. We have checked all the Hamiltonian forms given in Eqs.~\eqref{eq:2D_L_H}--\eqref{eq:coupled_QHO} and the obtained energy spectra are in excellent agreement with one another; see EM for more details. Two typical energy spectra are shown in Figs.~\ref{fig:energy_spectra}(a) and~\ref{fig:energy_spectra}(b) for two different values of $k_cl_B = 2$ and $\sqrt{5}\approx2.24$, respectively. For the chosen parameter $\Delta_c/\omega=-0.8$, in the absence of the coupling the first band at energy $E_1(0)=\hbar\omega/2$ corresponds to the lowest Landau level (LLL) and cavity vacuum, that is, $\ket{\Phi_{\ell,x_0}}\otimes\ket{n}=\ket{\Phi_{0,x_0}}\otimes\ket{0}$. This band is $m$-fold (here for the chosen parameters, 8-fold) degenerate, corresponding to the possible values of the guiding center $x_0$. The higher bands $\{E_2(0),E_3(0),E_4(0)\}$ in the shown spectra represent $\ket{\Phi_{0,x_0}}\otimes\ket{1}$ (LLL and one photon), $\ket{\Phi_{1,x_0}}\otimes\ket{0}$ (second LL and no photon), and $\ket{\Phi_{0,x_0}}\otimes\ket{2}$ (LLL and two photons), respectively.

%%--------Figure------------ 

By turning on the coupling $\eta\neq0$, the energy bands start to `disperse' due to the atom-photon mixing and some of the degeneracies are lifted~\cite{Zak1964Group, Capel1971, Rauh1974Degeneracy, Hofstadter1976Energy, Zhu1992Hofstadter}. The color map of the energy bands indicates the average photon number $\langle \hat{a}^\dag \hat{a} \rangle$ of the corresponding states. Note the absence of a sharp ``superradiant'' phase transition, as the system is composed of only a single atom.

In order to further elucidate the matter-field mixing, we decompose the eigenstates $\ket{\Psi_j}$ of the system into the bare uncoupled Landau and photon Fock bases,  
%\begin{align} \label{eq:Landau-polariton-decomp}
$\ket{\Psi_j}=\sum_{\ell,x_0,n} c_{\ell,x_0,n}^{(j)} \ket{\Phi_{\ell,x_0}}\otimes\ket{n}$. 
%\end{align}
The resultant probability amplitudes $|c_{\ell,x_0,n}^{(j)}|$ are illustrated in Figs.~\ref{fig:energy_spectra}(c) and~\ref{fig:energy_spectra}(d) for $k_cl_B = \sqrt{5}$ and two different coupling strengths $\eta/\omega=0.1$ and $1$, respectively. For the coupling $\eta=0.1\omega$, already adjacent bands with small energy gaps among them are mixed; notably, e.g., $\ket{\Phi_{0,x_0}}\otimes\ket{1}$ and $\ket{\Phi_{1,x_0}}\otimes\ket{0}$ in Fig.~\ref{fig:energy_spectra}(c). While the strong coupling $\eta=\omega$ mixes various matter-field states. This is a generalization of the Jaynes--Cummings-type polariton, hence the name ``\emph{Landau polaritons}''. Note that states within each band do not mix with each other, since the interaction Hamiltonian has no non-zero matrix element among them.

%=========================================================================
\emph{Landau-polaritons' properties.}---The Landau polaritons $\ket{\Psi_j}$ are pure states, confirmed by $\text{Tr}(\hat{\rho}^2)=1$ with $\hat{\rho}$ being the total matter-field density operator. Therefore, the von Neumann entropy $S_{\rm vN}=-\text{Tr}[\hat{\rho}_{\rm red}\log(\hat{\rho}_{\rm red})]$ of the reduced system $\hat{\rho}_{\rm red}=\text{Tr}_{a(b)}(\hat{\rho})$, with $\text{Tr}_{a(b)}$ being the partial trace over the photon (Landau) oscillator, can serve as a genuine measure of the matter-field entanglement. This is shown in the inset of Figs.~\ref{fig:energy_spectra}(a) and ~\ref{fig:energy_spectra}(b) for the corresponding lowest bands (note the partial degeneracy of some of the eight sub-bands), confirming light-matter entanglement, $S_{\rm vN}>0$. Note the saturation of the entanglement for some sub-bands in Fig.~\ref{fig:energy_spectra}(a) to one ebit, $S_{\rm vN}\simeq\log(2)$, signaling the emergence of Bell-like states within the infinite-dimensional Fock spaces.

To further highlight the consequences of the matter-field coupling, we calculate the variance $\{\sigma^2_{{\mathcal Q}_{O}}, \sigma^2_{{\mathcal P}_{O}} \}$ of quadrature operators $\hat{\mathcal Q}_{O}=(\hat{O}^\dag+\hat{O})/2$ and $\hat{\mathcal P}_{O}=i(\hat{O}^\dag-\hat{O})/2$ of both light and matter harmonic oscillators, $\hat{O}=\{\hat{a},\hat{b}\}$. The results are depicted in Fig.~\ref{fig:squeezing} as a function of $\eta$ for $k_cl_B=\sqrt{5}$. As can be seen, both matter and light are squeezed, because while for matter $\sigma^2_{{\mathcal Q}_{b}}<1/4$, for light $\sigma^2_{{\mathcal P}_{a}}<1/4$. The squeezing of the real quadrature of matter $\hat{\mathcal Q}_{b}$ can be understood intuitively in the original physical space: By increasing coupling, the light field is populated creating hence an optical lattice for the particle, which in turn localizes the atom along the $x$ direction around a potential minimum and hence squeezes its position~\cite{Fletcher2021}. On the other hand, the squeezing of the imaginary quadrature of the light $\hat{\mathcal P}_{a}$ is consistent with the Dicke superradiance~\cite{Emary2003Chaos}.

%%--------Figure------------ 

%=========================================================================
\emph{Nonequilibrium dynamics and fixed points.}---This system is intrinsically open and out of equilibrium as photons are continuously pumped to and lost from the system. Therefore, it is more appropriate to talk about dynamics and fixed points (i.e., steady states) of the system, rather than the eigenstates of the Hamiltonian. The nonequilibrium quantum dynamics of the system is governed by the master equation,
\begin{align} \label{eq:master-eq}
\frac{d}{dt}{\hat \rho}(t) = -\frac{i}{\hbar}[\hat{H}, \hat {\rho}]
+\kappa (2\hat{a}\hat {\rho}\hat{a}^\dag - \hat{a}^\dag\hat{a} \hat {\rho} - \hat {\rho}\hat{a}^\dag\hat{a}),
\end{align}
where $2\kappa$ is the photon decay rate. Once the time evolution of the density operator $\hat{\rho}(t)$ is obtained, one can calculate the expectation value of any observable $\hat{O}$ via $O \equiv \langle \hat{O} \rangle = \text{Tr}(\hat {\rho}\hat{O})$.

%%--------Figure------------ 

In order to obtain an \emph{intuitive} understanding of quantum dynamics of a complex system, it is often useful to also look at the semi-classical dynamics of the system, despite its limitations. To this end, we derive the semi-classical equations of motion, 
\begin{align} \label{eq:semi-classical-EoM}
\frac{da}{dt}&=i(\Delta_c+i\kappa)a 
-i \eta \cos\left[k_c\left( \sqrt{2}l_B b_{\rm R} + x_0 \right) \right],
\nonumber\\
\frac{db}{dt}&= -i\omega b 
+ i \sqrt{2} k_cl_B \eta a_{\rm R}
\sin\left[k_c \left( \sqrt{2}l_B b_{\rm R} + x_0 \right) \right],
\end{align}
by omitting all quantum correlations, i.e., $\mathcal{C}_{a^i\bar{a}^jb^k\bar{b}^l}\equiv\langle \hat{a}^i(\hat{a}^\dag)^j \hat{b}^k (\hat{b}^\dag)^l \rangle - a^i(a^*)^jb^k(b^*)^l=0$. For brevity, in Eq.~\eqref{eq:semi-classical-EoM} I have introduced $O_{\rm R}\equiv \text{Re}(O)$ for the real part of the amplitudes $\{a,b\}$. The fixed points are obtained by setting the time evolution of the amplitudes to zero, i.e., $da_{\rm ss}/dt=db_{\rm ss}/dt=0$. This leads to an equation for the steady-state cavity-field amplitude $a_{\rm ss}= \eta \cos[k_c(\sqrt{2}l_B b_{\rm ss,R} + x_0)]/ (\Delta_c+i\kappa)$ in terms of $b_{\rm ss,R}\equiv\text{Re}(b_{\rm ss})$, which can be subsequently substituted in the equation for the steady-state Landau-oscillator amplitude, yielding 
\begin{align} \label{eq:ss-b}
\left[\frac{\sqrt{2}\omega(\Delta_c^2+\kappa^2)}{\eta^2\Delta_ck_cl_B}\right]b_{\rm ss}
=
\sin\left[2k_c \left( \sqrt{2}l_B b_{\rm ss,R} + x_0 \right) \right].
\end{align}

This is a nonlinear implicit equation and cannot be solved analytically to yield an explicit solution for $b_{\rm ss}$. However, useful insight can be drawn from this equation. First, we note that the right-hand side of Eq.~\eqref{eq:ss-b} is always real, implying that in the steady state $\text{Im}(b_{\rm ss})=0$, i.e., $b_{\rm ss}=b_{\rm ss}^*=b_{\rm ss,R}$. Second, solutions (i.e., roots) of this equation can be conveniently visualized graphically. In particular, the right-hand side is an oscillatory function in the interval $[-1,1]$. Therefore, if the slope $s=\sqrt{2}\omega(\Delta_c^2+\kappa^2)/\eta^2\Delta_ck_cl_B$ of the linear term in the left-hand side is small enough (e.g., for large coupling strengths $\eta$), numerous roots can exist, thus entailing multistability in the system. Furthermore, in addition to the parameters in the slop $s$, the number and the value of the steady state(s) depend on the Landau-oscillator center $x_0$. The stability of the fixed points is then determined through the linear stability analysis as discussed in EM. 

We now examine the nonequilibrium quantum and semi-classical dynamics of the system, obtained numerically by integrating Eqs.~\eqref{eq:master-eq} and~\eqref{eq:semi-classical-EoM}, respectively. The results are presented in Figs.~\ref{fig:dynamics}(a)-\ref{fig:dynamics}(c) for $\eta=3\omega$,  $k_cl_B=1$, $\kappa=\omega$, and three different values of $x_0/l_B=\{-\pi/2,-3\pi/4,-\pi\}$, respectively. The column~(i) shows the dynamics of the system in the phase space of the average photon number $N_a=\langle \hat{a}^\dag \hat{a} \rangle$ vs.\ the average LL occupation number $N_b=\langle \hat{b}^\dag \hat{b} \rangle$. The stable (unstable) semi-classical fixed points of the system for the given parameters are denoted by the filled (crossed) dark-red circles; the filled dark-blue circles represent the quantum steady states obtained from the long-time dynamics. The black crosses indicate the initial states. As can be seen, the quantum dynamics (blue solid curves) and the quantum steady states deviate significantly from the semi-classical ones (red dashed curves). This is due to the quantum correlations and the statistical mixing $\text{Tr}(\hat{\rho}^2)<1$, shown in the second column (ii) of Fig.~\ref{fig:dynamics}, which have been omitted in the semi-classical treatment. Note that although for $x_0/l_B=-\pi$ the lowest order correlation $\mathcal{C}_{ab}$ shown in Fig.~\ref{fig:dynamics}(c-ii) vanishes, the higher order correlations are non-zero and significant; see EM.  

The steady-state atomic and photonic probability distributions, $P_\ell^{\rm at}$ and $P_n^{\rm ph}$, are shown in the (iii) and (iv) columns, respectively. As references, I also include the thermal $P_{\rm th}(m)=N^m/(N+1)^{m+1}$ and Poisson $P_{\rm P}(m)=e^{-N}N^m/m!$ distributions~\cite{GerryKnight2023} with the corresponding numerical steady-state atomic and photonic averages $N=\{N_{a, \rm ss},N_{b, \rm ss}\}$. The insets in panels (iii) and (iv) illustrate, respectively, the steady-state atomic and photonic phase-space $Q$ functions, with the filled red (crossed white) circles indicating the corresponding semi-classical stable (unstable) steady states, $b_{\rm ss}$ and $a_{\rm ss}$. As discussed earlier, for $k_cx_0=-\pi/2$ the system possesses a $\mathbb{Z}_2$ symmetry, and both the atomic and photonic $Q$ functions display this symmetry. For $k_cx_0=-3\pi/4$ and $-\pi$ on the other hand, the $Q$ functions have a single dominant contribution peaked around a semi-classical stable fixed point. In particular, for the atomic $Q$ function, this peak corresponds to the smallest semi-classical stable fixed point; note that for $k_cx_0=-\pi$ this fixed point is $b_{\rm ss}=0$ which can still lead to a non-zero cavity field $a_{\rm ss}\propto\cos[k_c(\sqrt{2}l_B b_{\rm ss,R} + x_0)]\neq0$. For the chosen system parameters, the photon distributions follow closely the Poisson distribution. However, the atomic distribution is more sensitive and diverse. While for $k_cx_0=-\pi/2$ it is a mixed-state Poisson distribution, for $k_cx_0=-3\pi/4$ and $-\pi$ it resembles more closely thermal and squeezed state distributions, respectively.         

Finally, we note that we have also found multiple coexisting quantum steady states with different characteristics in some parameter regimes. An example is given in EM.

%=========================================================================
\emph{Conclusions.}---We have uncovered an interesting nonlinearly coupled quantum harmonic oscillator model originating from  driven-dissipative, single-particle Landau polariton physics. This coupled harmonic oscillator model is formally very similar to tweezer-trapped atoms inside cavity~\cite{Ho2025, Zhang2024Cavity} and therefore can be implemented in many state-of-the-art cavity-QED experiments in near future~\cite{Brennecke2007, Klinder2015, Vaidya2018Tunable, Zhang2021, Helson2023}.  Furthermore, the proposed setup may find important applications in metrology and sensing owing to the light-matter entanglement and the two-mode squeezing~\cite{Lawrie2019}.

Filling these Landau-polariton states will allow one to access and explore many-body, topological aspects of the system~\cite{Smolka2014}. In particular, the key question is how the (quantized) transport properties of the atomic quantum-Hall system change under the position-dependent real cavity field and the formation of driven-dissipative softened Landau polaritons~\cite{Cardoso2026Cavity}. Another intriguing scenario to explore is whether cavity-mediated atomic interactions can be strong enough to push the system into the fractional quantum Hall regime, a major challenge in atomic simulators of quantum Hall effect~\cite{Nascimbene2025Simulating}. These issues will be considered in future works.

%=========================================================================
\begin{acknowledgments}
I acknowledge inspiring discussions with Helmut Ritsch, Manuele Landini, and Natalia Masalaeva. This research was funded in whole or in part by the Austrian Science Fund (FWF) [grant DOI: 10.55776/P35891]. For open access purposes, the author has applied a CC BY public copyright license to any author accepted manuscript version arising from this submission.
\end{acknowledgments}

%=========================================================================
\bibliography{Landau_polariton}

%=========================================================================
%\newpage
\appendix*
%\widetext
\onecolumngrid
\setcounter{equation}{0}
\setcounter{figure}{0}
\renewcommand{\theequation}{A\arabic{equation}}
\renewcommand{\thefigure}{A\arabic{figure}}

%=========================================================================
\section{End Matter}
\twocolumngrid 

\begin{figure}[t!]
\centering
\includegraphics [width=0.48\textwidth]{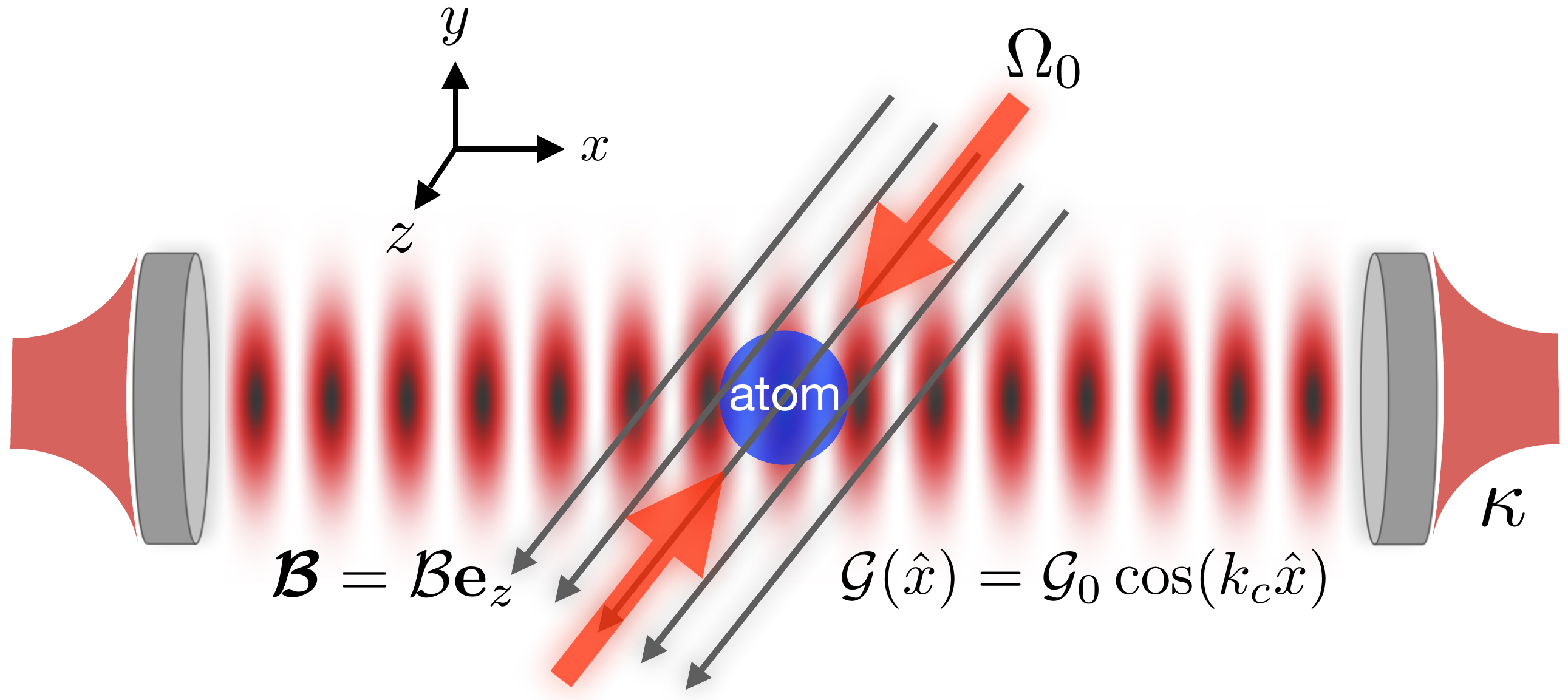}
\caption{Schematic sketch of the system. A single atom inside an optical cavity is box trapped in the $x$-$y$ plane and pierced by a transverse external synthetic magnetic field $\boldsymbol{\mathcal B}={\mathcal B}\mathbf{e}_z$ along the $z$ direction. The atom is coupled to a single off-resonant longitudinal mode of the cavity with the strength $\mathcal{G}(\hat{x})=\mathcal{G}_0\cos(k_c \hat{x})$, and is further driven in the transverse $z$ direction by an off-resonance standing-wave laser with the amplitude $\Omega_0$. The laser is closely red-detuned from the cavity resonance.
} 
\label{EM-fig:setup}
\end{figure}

%=========================================================================
\emph{Setup}.---The considered system was described in detail in the main text. In order to further elucidate the system, a sketch of the setup is presented here in Fig.~\ref{EM-fig:setup}.

%%--------Figure------------ 

\begin{figure}[t!]
\centering
\includegraphics [width=0.48\textwidth]{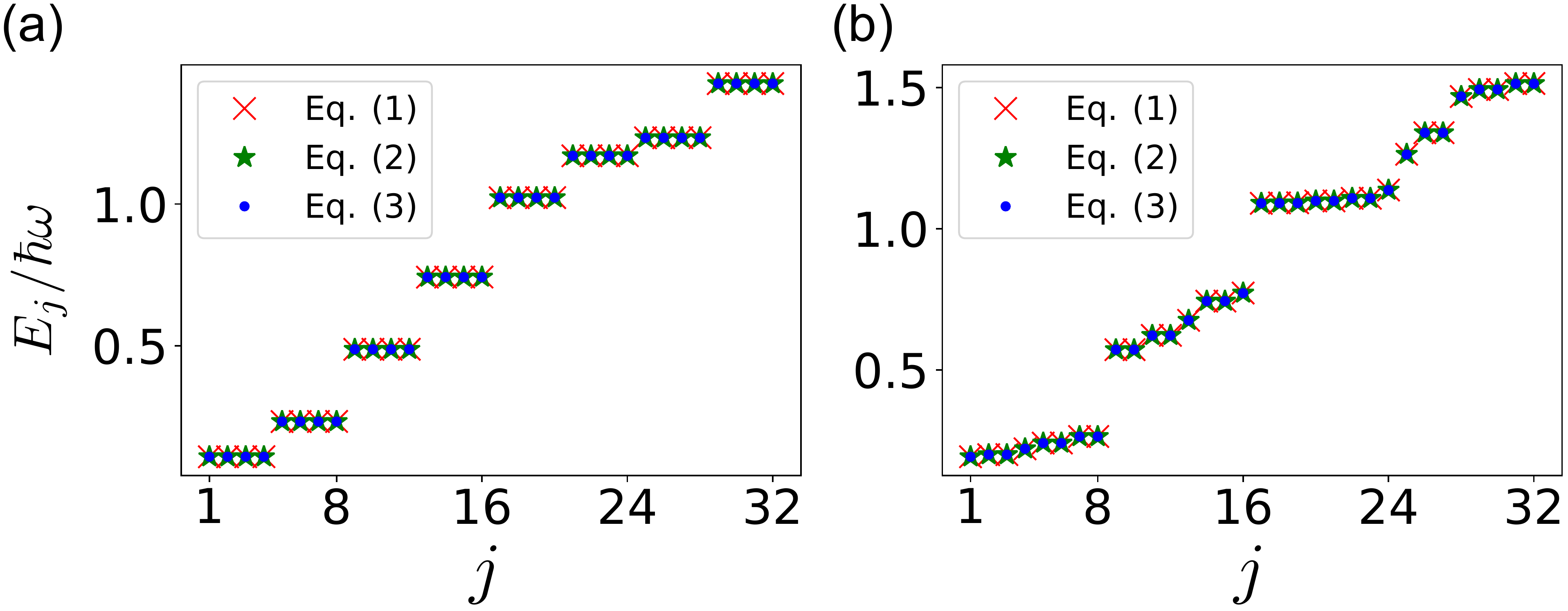}
\caption{Comparison of the energy spectrum $E_j$ calculated from the Hamiltonians~\eqref{eq:2D_L_H}, \eqref{eq:1D_L_H}, and~\eqref{eq:coupled_QHO} at the pump strength $\eta=\omega$ for two different values of $k_cl_B=2$ (a) and $\sqrt{5}\approx2.24$~(b), corresponding to cuts from Fig.~\ref{fig:energy_spectra} in the main text. In the two-dimensional case of Eq.~\eqref{eq:2D_L_H}, a system of size $20l_B\times8l_B$ with 60 (8) grid points along the $x$ ($y$) direction is chosen. The system size along the $y$ direction (i.e., $8l_B$) and its discretization ensure that for the periodic boundary condition they yield $x_0=\hbar k_y/\mathcal{B}=\pi l_B\{-1,-3/4,-1/2,\cdots,3/4\}$ as used in Fig.~\ref{fig:energy_spectra}. In the one-dimensional case of Eq.~\eqref{eq:1D_L_H}, 100 grid points are chosen for the same system size along the $x$ direction, $20l_B$. The photon cut-off is set to 15, 20, and 35, respectively, for diagonalizing the Hamiltonians~\eqref{eq:2D_L_H}, \eqref{eq:1D_L_H}, and~\eqref{eq:coupled_QHO}; we have checked the validity of these cut-offs a posteriori. The cut-off for the atomic oscillator is set to 20 in the coupled harmonic oscillator model of Eq.~\eqref{eq:coupled_QHO}, as only a few LLLs are involved for the number of bands and parameters shown. The other parameters are the same as Fig.~\ref{fig:energy_spectra}.
} 
\label{EM-fig:E-comparison}
\end{figure}

%=========================================================================
\emph{Validity and efficiency of the coupled harmonic oscillator description}.---In order to check the validity and the efficiency of the coupled harmonic oscillator description compared to the original forms in position space, I diagonalize all the forms of the Hamiltonian given in Eqs.~\eqref{eq:2D_L_H}, \eqref{eq:1D_L_H}, and~\eqref{eq:coupled_QHO} in the main text. The resultant thirty-two low-lying energies, corresponding to the four low-lying energy bands split by the atom-photon interaction, are illustrated in Fig.~\ref{EM-fig:E-comparison} for the coupling strength $\eta=\omega$ and two different values of $k_cl_B=2$~(a) and $\sqrt{5}\approx2.24$~(b). That is, these are cuts at $\eta=\omega$ of the energy spectra shown in Figs.~\ref{fig:energy_spectra}(a) and (b) in the main text, respectively. As can be seen, the energies calculated from the different forms of the Hamiltonian are in perfect agreement with one another. However, the coupled harmonic oscillator representation provides a much more efficient description as it has a more compact Hilbert space $\mathcal{H}=\mathcal{H}_a\otimes\mathcal{H}_b$ spanned by two Fock bases. This provides a convenient and practical framework especially in the strong coupling regime, where the number of involved and coupled states grow exponentially and the position space descriptions become unfeasible.

\begin{figure}[b!]
\centering
\includegraphics [width=0.48\textwidth]{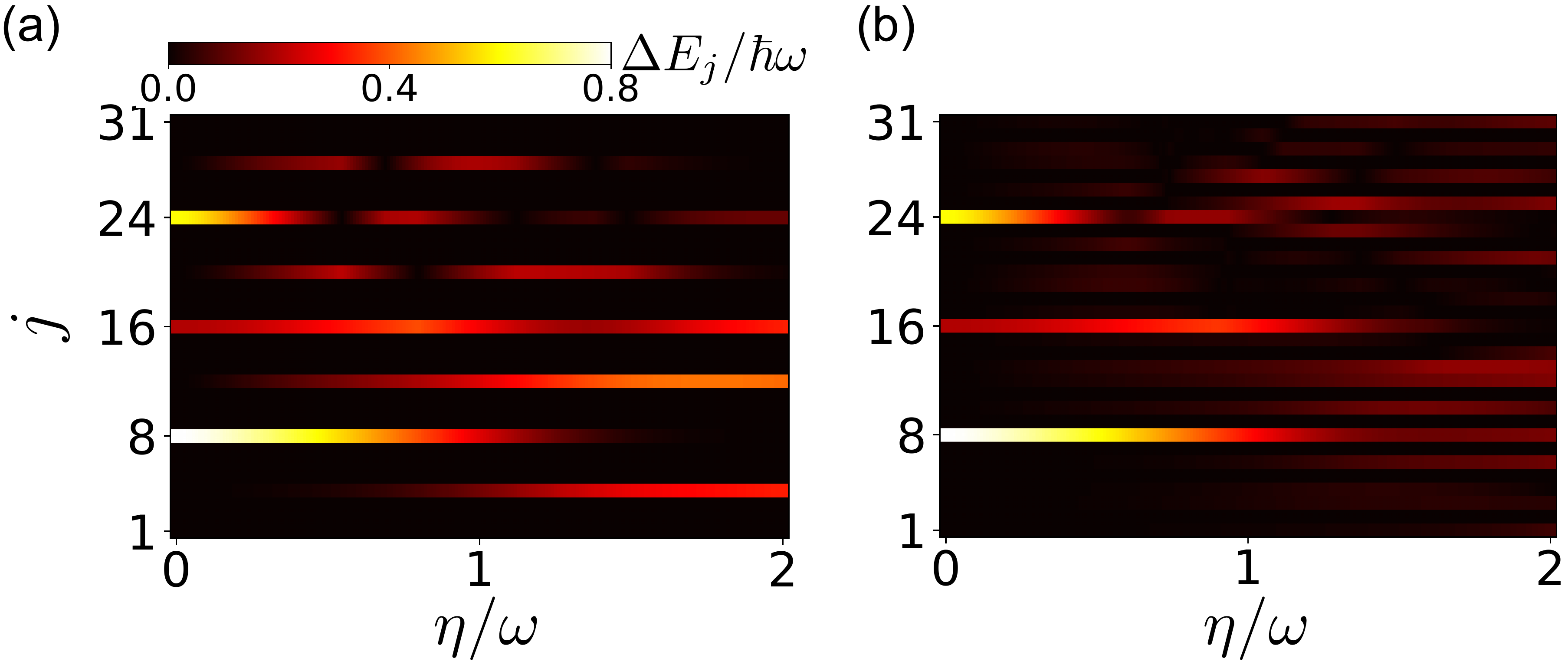}
\caption{The energy gap $\Delta E_j = E_{j+1} - E_j$ as a function of the atom-field coupling $\eta$ for two different values of $k_cl_B=2$~(a) and $\sqrt{5}\approx2.24$ (b), corresponding to the energy spectra shown in Figs.~\ref{fig:energy_spectra}(a) and \ref{fig:energy_spectra}(b), respectively. The other parameters are the same as Fig.~\ref{fig:energy_spectra}.
} 
\label{EM-fig:E-gaps}
\end{figure}

%=========================================================================
\emph{Energy gaps}.---In order to better visualize the splitting of the energy spectrum as a consequence of the light-matter interaction, here in Fig.~\ref{EM-fig:E-gaps} we display the energy gaps $\Delta E_j = E_{j+1} - E_j$, i.e., the energy difference between the adjacent levels, as a function of the atom-field coupling $\eta$ for two different values of $k_cl_B=2$~(a) and $\sqrt{5}\approx2.24$~(b). These figures correspond exactly to the energy spectra shown in Figs.~\ref{fig:energy_spectra}(a) and (b) in the main text, respectively. In the zero coupling limit, there are sharp, large gaps only every  eight states, corresponding to the degeneracy of the LLs (i.e., possible values of the guiding center $x_0$) in the numerics. For example, the first gap at $j=8$ represents the energy difference $\Delta E_8=0.8\hbar\omega$ between $\ket{\Phi_{0,x_0}}\otimes\ket{0}$ (LLL and the cavity vacuum) and $\ket{\Phi_{0,x_0}}\otimes\ket{1}$ (LLL and one photon); see also Fig.~\ref{fig:energy_spectra} and the main text. For $k_cl_B=\sqrt{5}$ shown in Fig.~\ref{EM-fig:E-gaps}(b) the energy spectrum exhibits more splitting, a common precursor to quantum chaos.

%%--------Figure------------ 

%%--------Figure------------ 

%========================================================================
\emph{Linear Stability Analysis}.---Let us begin by writing the semi-classical equations of motion, Eq.~\eqref{eq:semi-classical-EoM}, in the vector form $\partial_t\mathbf{v}=\mathbf{F}(\mathbf{v})$, where $\mathbf{v}=(a,a^*,b,b^*)^\intercal$ and $\mathbf{F}(\mathbf{v})$ is obtained from the right-hand side of Eq.~\eqref{eq:semi-classical-EoM}. However, it turns out that it is  convenient to work instead with real and imaginary parts of the mean-field amplitudes $\mathbf{u}=(a_{\rm R},a_{\rm I},b_{\rm R},b_{\rm I})^\intercal$, that is, 
\begin{align} \label{EM-eq:sc_EOM_compact}
\frac{\partial\mathbf{u}}{\partial t}=\mathbf{G}(\mathbf{u}),
\end{align}
where $\mathbf{G}(\mathbf{u})$ is related to $\mathbf{F}(\mathbf{v})$ via a coordinate transformation. 
The steady states $\partial_t\mathbf{u}_{\rm ss}=0$ are obtained by solving four non-linear coupled equations 
$\mathbf{G}(\mathbf{u}_{\rm ss})=0$,
leading to Eq.~\eqref{eq:ss-b} in the main text.

Expanding the right-hand side of Eq.~\eqref{EM-eq:sc_EOM_compact} using a Taylor series around the steady state $\mathbf{u}=\mathbf{u}_{\rm ss}+\delta\mathbf{u}$ up to first order in perturbations $\delta\mathbf{u}=(\delta a_{\rm R},\delta a_{\rm I}, \delta b_{\rm R}, \delta b_{\rm I})^\intercal$ and noting that $\partial_t\mathbf{u}_{\rm ss}=\mathbf{G}(\mathbf{u}_{\rm ss})=0$ yields a set of four coupled linearized equations,
\begin{align}
\frac{\partial \delta\mathbf{u}}{\partial t}=\mathbf{J}\delta\mathbf{u},
\end{align}
where $\mathbf{J}(\mathbf{u}_{\rm ss})= ({\partial\mathbf{G}(\mathbf{u})}/{\partial \mathbf{u}})\big|_{\mathbf{u}_{\rm ss}}$ is the Jacobian matrix,
\begin{align}
\mathbf{J} =
\begin{pmatrix}
-\kappa & -\Delta_c & 0 & 0 \\
\Delta_c & -\kappa & \sqrt{2}\eta \xi\sin\theta_{\rm ss} & 0 \\
0 & 0 & 0 & \omega \\
\sqrt{2}\eta \xi\sin\theta_{\rm ss} & 0 & -\omega+2\eta a_{\rm ss,R} \xi^2\cos\theta_{\rm ss} & 0
\end{pmatrix}.
\end{align}
Here, I have introduced $\theta_{\rm ss}\equiv k_c(\sqrt{2}l_B b_{\rm ss,R} + x_0)$ and $\xi\equiv k_cl_B$ for brevity.  A steady state $\mathbf{u}_{\rm ss}$ is stable if all eigenvalues of its corresponding Jacobian matrix $\mathbf{J}(\mathbf{u}_{\rm ss})$ have negative real parts. This is because then all fluctuations around that fixed point decay over time.

\begin{figure}[t!]
\centering
\includegraphics [width=0.48\textwidth]{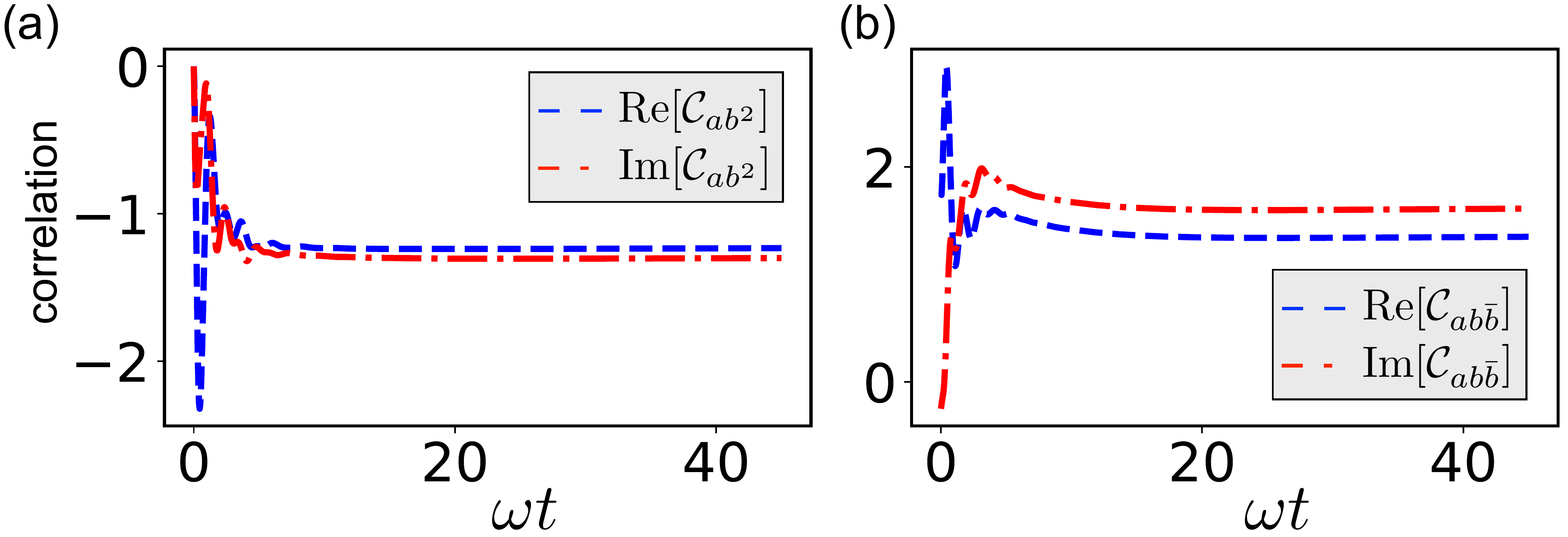}
\caption{The time evolution of the higher order correlations $\mathcal{C}_{ab^2}$ (a) and $\mathcal{C}_{ab\bar{b}}$ (b) for $k_cl_B=1$ and $x_0/l_B=-\pi$. The other parameters are the same as Fig.~\ref{fig:dynamics}.
} 
\label{EM-fig:corr-higher}
\end{figure}

%=========================================================================
\emph{Higher order correlations}.---As shown in Fig.~\ref{fig:dynamics}(c-ii) in the main text, the lowest order correlation $\mathcal{C}_{ab}=\langle \hat{a}\hat{b} \rangle - ab$ vanishes for $k_cx_0=-\pi$. However, this does not imply that the semi-classical description must be accurate, as the system is highly nonlinear. For example, I show in Figs.~\ref{EM-fig:corr-higher}(a) and~\ref{EM-fig:corr-higher}(b) the time evolution of the higher order correlations $\mathcal{C}_{ab^2}=\langle \hat{a}\hat{b}^2 \rangle - ab^2$ and $\mathcal{C}_{ab\bar{b}}=\langle \hat{a}\hat{b}\hat{b}^\dag \rangle - a|b|^2$, respectively. They both are nonzero and significant, highlighting the highly coupled nature of the system and the inadequacy of the semi-classical treatment in general.

%%--------Figure------------ 

Semi-classical description tends to become more accurate when all (lowest- and higher-order) correlations become negligible. For instance, for smaller $k_cl_B$ where the cosine coupling term in Eq.~\eqref{eq:coupled_QHO} can be truncated in some order the semi-classical and quantum dynamics agree better in general.

%=========================================================================
\emph{Quantum multistability}.---If the null space of the Liouvillian of an open quantum system consists of multiple, linearly independent vectors, then the system is multistable. Therefore, the long-time quantum dynamics and the quantum steady state depend on initial state, similar to the semi-classical dynamics. That is, for different initial states, the quantum dynamics can lead to different steady states. An example is displayed in Fig.~\ref{EM-fig:dynamics} for $\eta=3\omega$, $\kappa=\omega$, $k_cl_B=1$, $x_0/l_B=-\pi$, and two different initial states. In particular, Fig.~\ref{EM-fig:dynamics}(b) is just Fig.~\ref{fig:dynamics}(c) from the main text, reproduced here for the convenience of the comparison. As can be seen from Figs.~\ref{EM-fig:dynamics}(a-i) and \ref{EM-fig:dynamics}(b-i), the long-time quantum (as well as semi-classical) dynamics and steady states are different. Note as well that the properties of the steady states (in particular, the statistical mixing and the atomic distribution $P_\ell^{\rm at}$) are also very distinct, as shown in the other panels.

\onecolumngrid
\begin{center}
\includegraphics [width=0.98\textwidth]{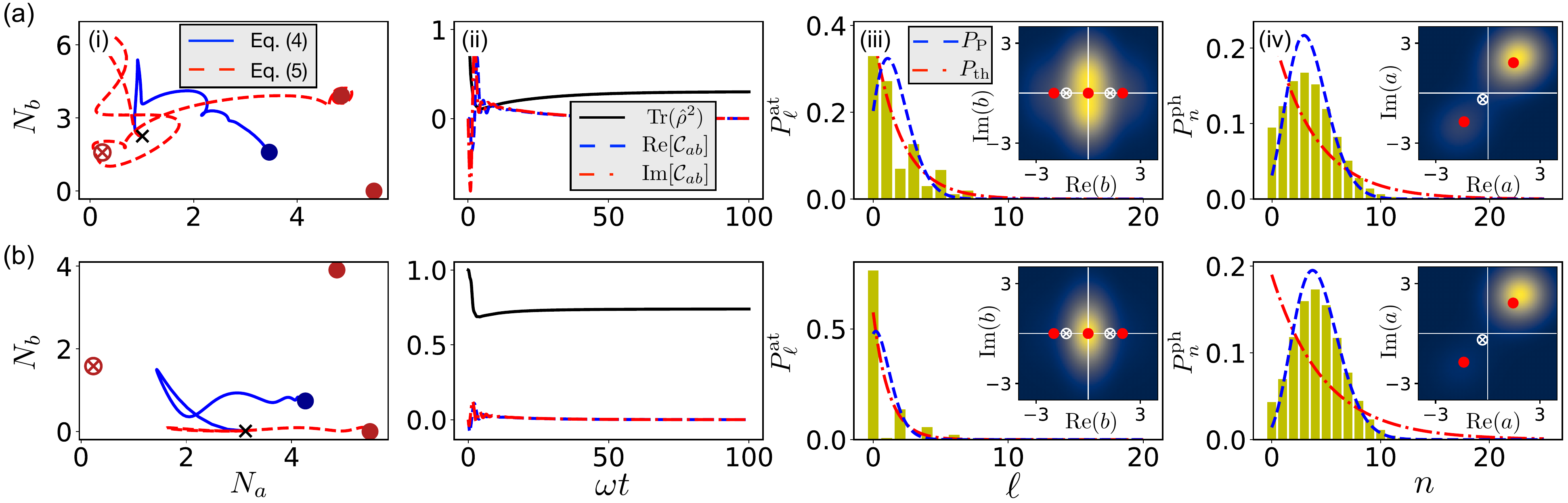}
\widefigcaption{The nonequilibrium dynamics of the system for $\eta=3\omega$, $\kappa=\omega$, $k_cl_B=1$, $x_0/l_B=-\pi$, and two different initial states: $\ket{1.5+0.05i}\otimes\ket{1-0.1i}$ (a) and $\ket{0.1+0.05i}\otimes\ket{1.75-0.25i}$ (b). The panel (b) is just Fig.~\ref{fig:dynamics}(c) from the main text, included here for the convenience of the comparison. The long-time quantum (as well as the semi-classical) dynamics and steady states dependent on the initial states, owing to the multistability.}
\label{EM-fig:dynamics}
\end{center}
\twocolumngrid

%\begin{figure*}[b!]
%\centering
%\includegraphics [width=0.97\textwidth]{Fig_SM_dynamics.pdf}
%\caption{The nonequilibrium dynamics of the system for $\eta=3\omega$, $\kappa=\omega$, $k_cl_B=1$, $x_0/l_B=-\pi$, and two different initial states: $\ket{1.5+0.05i}\otimes\ket{1-0.1i}$ (a) and $\ket{0.1+0.05i}\otimes\ket{1.75-0.25i}$ (b). The panel (b) is just Fig.~\ref{fig:dynamics}(c) from the main text, included here for the convenience of the comparison. The long-time quantum (as well as the semi-classical) dynamics and steady states dependent on the initial states, owing to the multistability.}
%\label{EM-fig:dynamics}
%\end{figure*}

\end{document}